%% file: Template_ISBI_latex.tex
\documentclass{article}
\usepackage{amsmath,graphicx}

\usepackage[preprint]{spconf}
\toappear{To appear in {\it ISBI2025,
   April 14-17, 2025, Houston, USA}}
\usepackage[acronym,nohypertypes={acronym,notation}]{glossaries}

\usepackage{enumitem}
\setlist{nosep, leftmargin=14pt}
\usepackage{float}
\usepackage{mwe} % to get dummy images

\usepackage{booktabs}
\usepackage{multicol}
\usepackage{multirow}

\title{Improving Quality Control of MRI Images Using Synthetic Motion Data}
\name{\shortstack{C Bricout\thanks{
\hspace{-15pt}$^*$Equal senior author contribution. Correspondence to: 
\tt\small bricout.charles@outlook.com
}$^{1}$,
K Cho$^{2}$, M Harms$^{3}$, O Pasternak$^{2}$, C Bearden$^{4}$, PD McGorry$^{5}$, RS Kahn$^{6}$,\protect\\ JM Kane$^{7}$, \textit{B Nelson}$^{5}$\textit{, SW Woods}$^{8}$\textit{, ME Shenton}$^{2}$\textit{, S Bouix}$^{\ast 1}$\textit{, S Ebrahimi Kahou}$^{\ast 9, 10}$
}}

\address{
  $^{1}$École de technologie supérieure, 
  $^{2}$Mass General Brigham,
  $^{3}$Washington University,
  $^{4}$UCLA,
  $^{5}$Orygen,\\
  $^{6}$Mt Sinai,
  $^{7}$Northwell Health,
  $^{8}$Yale,
  $^{9}$University of Calgary,
  $^{10}$Canada CIFAR AI Chair/Mila
}
\glsdisablehyper
\begin{document}
%\ninept
%
\input{acronym}

\maketitle

\begin{abstract}
    MRI quality control (QC) is challenging due to unbalanced and limited datasets, as well as subjective scoring, which hinder the development of reliable automated QC systems. To address these issues, we introduce an approach that pretrains a model on synthetically generated motion artifacts before applying transfer learning for QC classification. This method not only improves the accuracy in identifying poor-quality scans but also reduces training time and resource requirements compared to training from scratch. By leveraging synthetic data, we provide a more robust and resource-efficient solution for QC automation in MRI, paving the way for broader adoption in diverse research settings.
\end{abstract}
\section{Introduction}
\label{sec:intro}
\Gls{mri} has become an invaluable tool to study the brain anatomy. This modality is however subject to multiple sources of artifacts \cite{bernstein_imaging_2006}, the most frequent being motion related artifacts. For anatomical \gls{t1w} imaging, even a small amount of motion can bias the estimation of cortical thickness \cite{alexander-bloch_subtle_2016}. 
While there are in-scanner methods to estimate and correct motion artifact, these methods are not broadly used \cite{tisdall_volumetric_2011, pollak_quantifying_2023}. 
Deep Learning has recently shown promise for this task with 2D and 3D motion classification (e.g., \cite{kustner_automated_2018, roecher_motion_2024, fantini_automatic_2021, bottani_automatic_2022}) and regression of \gls{rms} deviation for transforms \cite{pollak_estimating_2023}. 
Unfortunately, most datasets are relatively small, are not labeled for presence of motion artifacts, and are very unbalanced (ideally only a few scans are affected by severe motion), making training a motion classifier difficult.
One strategy to overcome this problem is the use of synthetic artifact generation. 
For example, in \cite{loizillon_automatic_2024}, a network is trained on synthetic data to perform binary classification and transferred to perform pass or fail classification of clinical data. 
In \cite{mohebbian_classifying_2021}, artifacts were synthetically produced on a wider range of motion and categorized into 5 bins, a 2D model was then trained for this synthetic classification task. 
In this paper, we improve upon these models by using a regression task estimating a \textit{continuous scalar motion score} \cite{jenkinson_measuring_nodate,pollak_estimating_2023}
as a pretraining objective instead of a binary classification \cite{loizillon_automatic_2024}. We train the model on synthetically generated motion data and further use transfer learning to perform \gls{qc} classification of volumes from the \gls{ampscz} dataset. We show this approach provides better results than training from scratch, given highly unbalanced dataset.
% \begin{table}[H]
%     \centering
%     \begin{tabular}{l  c c}
%          \toprule
%          & Synthetic Motion Generation  & QC task  \\
%          \midrule
%        AMPCSZ  & 319  & 197\\
%        HCPEP & 143 & 0  \\
%          \midrule
%        Total & 462 & 197 \\
%        \bottomrule
%     \end{tabular}
%     \caption{Final number of volume for each task, after filtering.}
%     \label{tab:num_volumes}
% \end{table}
\section{Materials and Methods}
\label{sec:methodology}

\noindent\textbf{Dataset:}
We use two datasets with manual \gls{qc} scoring. First, the \gls{hcpep} includes 390 3T \gls{t1w} \gls{mri} acquired at 3 different sites. Each volume receives a \gls{qc} score on a 4-point scale (1=poor, 2=fair, 3=good, 4=excellent) \cite{jacobs_introduction_2024}. This dataset was used exclusively for synthetic motion generation and training.
The \gls{ampscz} \cite{wannan_accelerating_2024} dataset includes 1,048 3T \gls{t1w} scans acquired across 33 sites and \gls{qc}-scored using the same 4-point scale.
We applied Clinica's \texttt{t1-linear} pipeline \cite{routier_clinica_2021} to correct for bias field and align to the MNI152 template using affine registration. 
This dataset was split into 2 subsets, one for synthetic motion generation and training and one for \gls{qc} classification.
The split was performed based on the site to ensure that our downstream network are always trained on unseen sites. 26 sites were reserved for synthetic data and 7 for \gls{qc}. 
Synthetic motion generation was performed on data with a \gls{qc} score of 4 and no mention of motion in the \gls{qc} comments to avoid adding motion to a volume already displaying significant movement. We used 319 \gls{ampscz} and 143 \gls{hcpep} volumes for synthetic data generation and 378 \gls{ampscz} volumes for transfer learning.

\begin{figure}[h]
    \centering
    \includegraphics[width=1\linewidth]{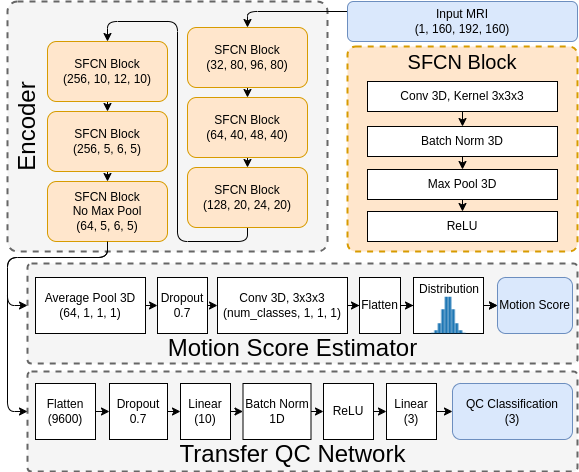}
    \caption{\gls{sfcn} architecture for training and transfer learning}
    \label{fig:model:sfcn}
\end{figure}

\noindent\textbf{Model Architectures:}
The architecture used to predict synthetic motion was the \gls{sfcn}  (Figure \ref{fig:model:sfcn}) used in \cite{pollak_estimating_2023}. 
For regression tasks (estimating the scalar motion score), we used the same strategy as \cite{pollak_estimating_2023}: define a range of values, discretize this range into 50 bins, learn a distribution over those bins, and reconstruct the final regression value by computing the sum of all bins probability multiplied by their center.
For transfer learning (estimating manual QC classes), we define a multi-layer perceptron to estimate classification based on \gls{sfcn} embeddings consisting of two linear layers with batch normalization and ReLU activation on the first layer.
Models were implemented using Pytorch \cite{paszke_pytorch_2019}, Lightning \cite{falcon_pytorchlightningpytorch-lightning_2020} and MONAI \cite{cardoso_monai_2022}.

\begin{figure}[h]
    \centering
    \includegraphics[width=0.75\linewidth]{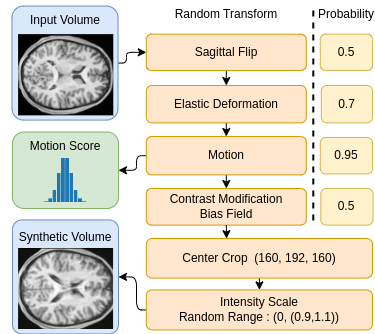}
    \caption{Pipeline to generate volumes with synthetic artifacts}
    \label{fig:synthpipe}
\end{figure}

\noindent\textbf{Pretraining for scalar motion estimation:}
To create our synthetic data, we designed a pipeline (Figure \ref{fig:synthpipe}) to randomly apply synthetic motion using TorchIO \cite{perez-garcia_torchio_2021}.
This transformation samples $N$ affine matrices representing subject motion and concatenate their k-space in a final, corrupted, k-space ~\cite{shaw_mri_2019}.
To expose our network to more variety, we added other random transforms: elastic deformation, bias field, contrast, flip on the sagittal plane and random scaling.
Finally, we cropped our volume to a ROI of (160, 192, 160).  
We split the data for train, validation and test before sampling to avoid subject data leakage. 
We sampled 300 forward passes through this random pipeline for each available volume, resulting in $110100$ volumes for training, $14100$ volumes for validation and $13801$ volumes for testing. 
The ground truth motion score was obtained by computing the \gls{rms} deviation from all affine matrices used to generate motion corrupted volumes \cite{jenkinson_measuring_nodate}. 
This provides a summary measure of the amount of subject motion synthetically generated and is the objective of the training.

\begin{table}[H]
    \centering
 \fontsize{9pt}{9pt}\selectfont
    \begin{tabular}{c c c c}
    \toprule
        Model & \multicolumn{2}{c}{SFCN Model} & Transfer QC\\
        &  Encoder &  Classifier & \\
        \midrule
        \#Parameters & 2,950,336 & 2,600 & 96,083\\
        \bottomrule
    \end{tabular}
    \caption{Number of parameters for each model used.}
    \label{tab:num_param}
\end{table}

Following \cite{pollak_estimating_2023}, we convert the scalar motion score to a discrete distribution over 50 bins, representing the range $[-0.8,4.8]$. 
We use the KL-Divergence between target and prediction motion distributions as our loss. 
We trained using AdamW with a learning rate of $2\times10^{-5}$ and a weight decay of $0.05$, we used a scheduler to decrease our  learning rate  by a factor of $0.6$ when getting no improvement for 5 epochs. 
We used a batch size of 24 per GPU and trained with four A100 GPU reaching an effective batch size of 96 volumes. 
We stopped training when the validation loss stopped progressing for more than 15 epochs, and we reported the model with the best validation coefficient ($R^2$).

\noindent\textbf{Transfer Learning for \gls{qc} classification:}
To assess the efficacy of pretraining on a synthetic motion estimation task to perform \gls{qc} classification, we compared this transfer learning task to training a model from scratch using \gls{ampscz} data. 
Given the scarcity of poor(=1) and fair(=2) scores in this dataset, we merged those two categories and trained the models to predict a 3-point scale. (Table \ref{tab:score_split})

\begin{table}[h]
    \centering
    \fontsize{9pt}{9pt}\selectfont
    \begin{tabular}{l c c c}
    \toprule
        QC Score & 1/2 & 3 & 4 \\
        \midrule
        Train &7 &38&70 \\
        Validation & 1 & 12& 26\\
        Test & 9 & 90 & 125\\
        \bottomrule
    \end{tabular}
    \caption{Data split for \gls{qc} classification task}
    \label{tab:score_split}
\end{table}

For both approaches (training from scratch or training on our pretrained model embeddings w/ frozen weight), hyperparameters were optimized using Ray Tune. 
Both models used a batch size of 12. 
The scratch model had a learning rate of $3\times10^{-6}$, dropout of 0.68, and weight decay of 0.06, while the transfer model used a learning rate of $ 5\times10^{-4}$, dropout of 0.7, and weight decay of 0.05. 
For scratch, we stopped training when the validation loss stopped diminishing for more than 100 epochs. For transfer learning, we limited the max number of epoch to 50. 
For both models, we reported the one with the best validation balanced accuracy. Every training iteration ran on one A100 GPU. 

\section{Results}
\label{sec:results}

\begin{figure}[t]
        \centering
        \includegraphics[width=0.7\linewidth]{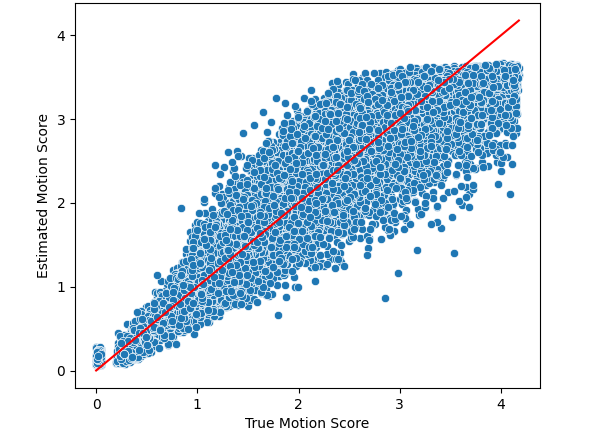}
        \caption{Calibration curve on validation dataset for the best epoch. Best R² correlation: 0.89}
        \label{fig:calibration}
\end{figure}

\noindent\textbf{Pretraining motion estimation task:}
Our model trained for 25h29m and reached a maximum validation $R^2$ of 0.89. 
The models performed similarly on the test set, indicating good generalization to unseen data (Figure \ref{fig:calibration}).
 \begin{figure}[h]
         \centering
         \includegraphics[width=0.9\linewidth]{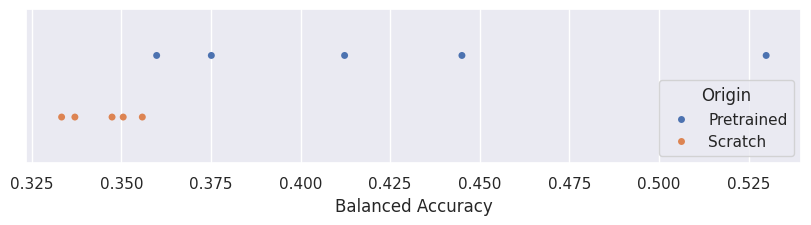}
         \caption{Comparison of test accuracy for each transfer model with pretrained task and models trained from scratch}
         \label{fig:transf:compar}
\end{figure}

\noindent\textbf{QC classification:}
We trained our models with 5 different random seeds to account for variations. We can see that transfer learning always outperforms training from scratch (Figure \ref{fig:transf:compar}). Table \ref{tab:acc_results} shows that models trained from scratch are unable to classify the first class (poor/fair quality). Looking at the F1 score ($2\cdot\frac{\text{precision}\cdot\text{recall}}{\text{precision}+\text{recall}}$) the models trained from scratch behave almost like a majority voter, mostly predicting the class 2, whereas every seed of transfer learning predicted some volumes for each class. 
\begin{table}[h]
    \centering
    \fontsize{9pt}{9pt}\selectfont
    \begin{tabular}{llrrrr}
        \toprule
        & &Balanced & \multicolumn{3}{c}{F1 Score by Class}\\
        Model & Base &  Accuracy & 1/2 & 3 & 4 \\
        \midrule
        \multirow[c]{2}{*}{SFCN}         & Pretrained & \bfseries 0.41 & \bfseries 0.13 & \bfseries 0.34 & 0.68  \\
         & Scratch & 0.35 & 0.00 & 0.20 & \bfseries 0.70 \\
        \bottomrule
    \end{tabular}
    \caption{Median result for scratch and transfer learning.}
    \label{tab:acc_results}
\end{table}
When looking at accuracy, using the pretrained model clearly outperforms training from scratch. Additionally, it is more resource efficient. Using transfer learning on our pretrained network does not require access to powerful GPUs to be trained and takes significantly less time to train. However, if we include the pretraining cost, it would necessitate 33 uses for the overall approach to become more time-efficient.

\begin{table}[h]
    \centering
%\resizebox{0.5\linewidth}{!}{%
    \fontsize{9pt}{9pt}\selectfont
    \begin{tabular}{lccc}
        \toprule
        & Max GPU & Max GPU  & Duration  \\
        &  Ram (GB) & Power (Watt) & (hh:mm:ss) \\
        \midrule
        Pretrain & 39.54 & 403.94 & 25:28:56 \\
        \midrule
        Transfer & 2.50 & 76.79 & 00:03:02 \\
        Scratch & 37.41 & 373.29 & 00:19:29 \\
        Decreased(\%) & 93.32\% & 79.43\% & 94.00\% \\
        \bottomrule
    \end{tabular}
    \caption{Differences in resource usage between both settings}
    \label{tab:res_results}
\end{table}

\section{Discussion}
\label{sec:discussion}
Our pretrained model is able to accurately predict the scalar motion score from synthetically motion-corrupted \gls{t1w} data. 
Furthermore, models trained to quantify motion appear to learn meaningful embeddings that can be leveraged to perform \gls{qc} classification of real \gls{mri} data. 
We also show that training on synthetic data can help with highly unbalanced datasets, where standard deep learning approaches fail to learn classes with few samples. 
%Our method shows more robustness. 
Overall, using transfer learning is more data efficient than training from scratch, and a model can be pretrained with less than 500 original samples by using synthetic artifact generation. 
%It is also important to note that for the same time invested in fine-tuning hyperparameters, using transfer learning seems a better approach.
Transfer learning is also significantly more resource efficient as the transfer task can be trained on a small GPU.

In terms of limitations, our best model has relatively low performance. 
This could be improved by pretraining on multiple objectives reflecting different kinds of artifacts. Data augmentation strategies could also be applied. We should also mention that our method has only been tested for 3T T1w data, but could easily be extended to other modalities.
Finally, it will be important to test this method on multiple \gls{qc} datasets to get a better understanding of its generalizability. 
To the best of our knowledge, this is the first attempt to learn regression for subject motion with synthetically generated artifacts and to transfer this motion-specific knowledge to a more general \gls{qc} classification task. 

More broadly, our results show that synthetic data can help with extreme data scarcity and that pretraining a model on a objective quantitative task before fine tuning on a subjective qualitative task is a promising approach. 
%Future research should expand this method to other artifacts and test whether they lead to improved subjective QC score prediction.
%It would also be interesting to investigate the usability of the predicted motion score for statistical correction of anatomical measurements.
\section{Conclusion}
\label{sec:conclusion}
\gls{mri} research needs automated \gls{qc} as the number of subjects per study keeps increasing.
Available \gls{qc} datasets are highly unbalanced and use different and subjective scales, making it hard to train one model that fits all scenarios.
In this paper, we present an approach to reduce the difficulty of training with unbalanced datasets and in a data-scarce setting for \gls{mri} \gls{qc}. 
Our results demonstrate the advantage of pretraining a model on synthetically generated motion artifacts before training for \gls{qc} classification. 
Even though motion is fully simulated, and our downstream task is different, using transfer learning on a pretrained model yields better results than training from scratch. 
Furthermore, using transfer learning consumes less resources and requires less time. This pretraining strategy has the potential to allow research teams to automate their own specific and subjective \gls{qc} pipeline and, eventually, render \gls{qc} scales less subjective.   
% \begin{itemize}
%     \item Future Works, possibility one : Generate data using other dataset than AMPSCZ and HCPEP to get two possible datasets with clear scoring and larger amount of data (scoring is needed to be sure volumes are clear before artifact generation, maybe high amount of data can compensate for potentially noisy labels). Design experiment with multiple objective for multiple artifacts to learn representation on overall quality
%     \item Future Works, possibility two : Study the value of our learned regression in regard of WM estimation/statistical correction and potential vNav scores on HCP aging and development
% \end{itemize}
\section{Ethics and Acknowledgements}
\label{sec:ethics}
This research used retrospective human subject data made available by the NIMH Data Archives. Approval was granted by the Research Ethics Committee of École de technologie supérieure. 
We thank the Accelerating Medicines Partnership® Schizophrenia (AMP® SCZ), a public-private partnership managed by the Foundation for the National Institutes of Health and supported by contributions from the AMP SCZ public and private partners, which include NIMH (U24MH124629, U01MH124631, and U01MH124639) and Wellcome (220664/Z/20/Z and 220664/A/20/Z).
This research is also supported by the Canada Research Chairs Program (SB), CIFAR (SEK) and DRAC.

\bibliographystyle{IEEEbib}
\bibliography{refs}

\end{document}

%% file: acronym.tex
\newacronym{mri}{MRI}{Magnetic Resonance Imaging}
\newacronym{hcpep}{HCPEP}{Human Connectome Project Early Psychosis}
\newacronym{ampscz}{AMP SCZ}{Accelerating Medecine Partnership Scizophrenia}
\newacronym{rms}{RMS}{Root Mean Square}
\newacronym{qc}{QC}{Quality Control}
\newacronym{sfcn}{SFCN}{Simple Fully Convolutional Network}
\newacronym{t1w}{T1w}{T1-weighted}